\documentclass[aps,prb,twocolumn,superscriptaddress,floatfix,letterpaper]{revtex4-1}
\usepackage[inline]{enumitem}
\usepackage{tikz}
\usepackage[export]{adjustbox}
\usepackage{float}
\usepackage[normalem]{ulem}

\pdfoutput=1

\def\arraystretch{1.25} 

\newcommand{\nn}{\nonumber}

\renewcommand{\tilde}{\widetilde}
\renewcommand{\bar}{\overline}

\renewcommand{\hat}[1]{\widehat{#1}}
\newcommand{\bchi}{\bar{\chi}}

\newcolumntype{L}{>{$}l<{$}}
\newcolumntype{R}{>{$}r<{$}}
\newcolumntype{C}{>{$}c<{$}}

\newcommand{\calH}{\mathcal{H}}

\newcommand{\calO}{\mathcal{O}}
\newcommand{\calT}{\mathcal{T}}
\newcommand{\calS}{\mathcal{S}}
\newcommand{\bpsi}{\bar \psi}
\newcommand{\ZZ}{\mathbb{Z}}

\newcommand{\fh}{\frac{1}{2}}

\begin{document}

\begin{titlepage}
\title{A continuous topological phase transition between\\
 two 1D anti-ferromagnetic spin-1 boson superfluids with the same symmetry
}


\author{Wenjie Ji}
\affiliation{Department of Physics, Massachusetts Institute of
Technology, Cambridge, Massachusetts 02139, USA}
\author{Xiao-Gang Wen}
\affiliation{Department of Physics, Massachusetts Institute of
Technology, Cambridge, Massachusetts 02139, USA}

\begin{abstract} 
Spin-1 bosons on a 1-dimensional chain, at incommensurate filling with anti-ferromagnetic spin
interaction between neighboring bosons, may form a spin-1 boson condensed state
that contains both gapless charge and spin excitations.  We argue that the
spin-1 boson condensed state is unstable, and will become one of two
superfluids by opening a spin gap.  One superfluid must have a spin-1 ground
state on a ring if it contains an odd number of bosons and has no degenerate
states at the chain end.  The other superfluid has a spin-0 ground state on a
ring for any numbers of bosons and has a spin-$\fh$ degeneracy at the chain end.
The two superfluids have the same symmetry and only differ by a spin-$SO(3)$
symmetry protected topological order.  Although Landau theory forbids
a continuous phase transition between two phases with the same symmetry, the
phase transition between the two superfluids can be generically continuous,
which is described by a conformal field theory (CFT) $su(2)_2\oplus u(1)_4 \oplus
\bar{su(2)}_2\oplus \bar{u(1)}_4$. Such a CFT has a spin fractionalization: 
spin-1 excitation can decay into a spin-$\fh$ right mover and a spin-$\fh$ left
mover. We determine the critical theory by solving the partition function based on emergent symmetries and modular invariance condition of CFTs.

\end{abstract}


\maketitle

\end{titlepage}


\noindent

\textbf{Introduction}: 1+1D spin-1 charge-2 boson system at incommensurate
filling has $SO(3)\times U(1)\times U_t(1)$ symmetry, where $U_t(1)$ is the
translation symmetry. Besides the boson condensing state that has gapless
spin-1 charge-2 excitations, anti-ferromagnetic spin interactions can lead to
superfluid states. The superfluids have fully gapped spin
excitations\cite{essler2009spin} and gapless charge excitations. (For integral
filling, see \Ref{zhou2003spin}.)  Under the protection of $SO(3)\times
U(1)\times U_t(1)$, there can be two such kinds of superfluids:\\
(1) two spin-1 bosons form a spin-singlet bound state and the spin-0 charge-4
bosons form a superfluid, which is referred as non-topological superfluid
(nTSF).\\  
(2) In some sense, the spin-1 bosons fractionalize into charge-1 spin-$\fh$'s.
The spin-$\fh$'s form singlet pairs and then the spin-0 charge-2 pairs form a
superfluid.  Or more precisely, the spin part is analogous to the AKLT state in
a spin-1 chain\cite{H8364,AKL8877}. The resulting superfluid is referred as
topological superfluid (TSF).\\
For a system on a ring with $N$ charge-2 spin-1 bosons, when $N$ is even, both
the nTSF and TSF has a spin-0 ground state. However, when $N$ is odd, the
ground state of the nTSF has spin-1, (resembling the dimerized phase in spin-1
antiferromagnetic chain,) while that of the TSF is still a spin-0 state.
Moreover, for a system with open ends and $N$ is even, only the TSF has
emergent a spin-$\fh$ at each end. 


Conventional Landau theory forbids two phases with the same symmetry to have a
continuous phase transition between them. In this paper, we show that there is
a consistent theory to describe a continuous phase transition between the
topological and the non-topological superfluids. (Some examples of topological
phase transitions that do not change the symmetry can be found in
\Ref{WW9301,CFW9349,W0050,SVB0490,senthil2006competing,ran2006continuous,sachdev2010quantum}.)
The critical point is described by a conformal field theory (CFT) built from
$u(1)_M$ and $su(2)_k$ current algebras\cite{francesco2012conformal}. We find
the phase boundary of the two superfluids may allow more than one CFT that
share the same lattice symmetry.  Among them, only one set of CFTs, namely
$u(1)_M\oplus su(2)_2\oplus \bar{u(1)}_M\oplus \bar{su(2)}_2$ that are
consistent candidate of critical theory with a single symmetric relevant
operator (and smallest central charge). We argue that this theory describes the
spin-1 boson condensed state with anti-ferromagnetic interaction. When
perturbing the  state with the only symmetric relevant operator, the system
flows to either one of the above two superfluid phases. Meanwhile, we hope to
provide a new way to find consistent  symmetric critical theories based on
modular invariance of CFTs and anomaly matching condition of symmetry charges.
We also proposed a bosonic model that realizes the topological superfluid
phase. 

\noindent\textbf{Topological superfluid}: We propose to find the TSF phase in
the doped t-J model composed of hard-core spin-1 bosons. The Hamiltonian of one
such model could be
\begin{align}
H=&-t\sum_{j,\sigma}  (b_{j,\sigma}^\dagger b_{j+1,\sigma}+h.c.)]+U\sum_{j} n_j (n_j-1)\nn\\
&+\sum_j \left[J_1\vec S_j\cdot \vec S_{j+1}-J_2 (\vec S_j\cdot \vec S_{j+1})^2\right]
\label{ham}
\end{align}
where $\sigma=\pm 1,0$ is the spin index, $\vec S_j$ is the spin on site $j$
and $n_j=\sum_{\sigma=\pm 1,0} b_{j,\sigma}^\dagger b_{j,\sigma}$ is the boson
density on site $j$, which we restrict to be $0$ or $1$.  When the boson number
$N$ and site number $N_L$ are equal $N=N_L$, the above system is in the AKLT
state for small $J_2$ and is in the dimmer state for a larger $J_2>J_1>0$.
After we add some small doping, $N < N_L$ and $\frac{N_L}{N} \approx 1$, we
expect the AKLT state to become the TSF state and the dimmer state   to become
the nTSF state.\footnote{One may need to add additional term $t_2\sum_{j,\sigma}b_{j,\sigma}^\dagger b_{j+2,\sigma}$ to find the nTSF.}


\noindent
\textbf{Partition functions of $U(1)\times SO(3)$ symmetric critical theories}:
In general, it is difficult to find the critical point of strongly interacting
models such as (\ref{ham}). In this paper, we provide a
procedure to determine the low energy theories of possible critical points, by
identifying partition functions, based on current algebra and modular
invariance of a conformal field theory. Physical data of a symmetric critical
point is all low energy excitations, their energies as a function of momentum,
together with their charges under the symmetry. The 1+1d critical point has the
nice feature that the low energy gapless excitations can be described by a
conformal field theory. When the critical point is between phases under
symmetry protection, the CFT has extended current algebra. One famous example
with explicit Lagrangians is Wess-Zumino-Witten (WZW)
models.\cite{witten1984non,knizhnik1984current} Traditional treatments of the
critical theory start with first mapping the spin model to a free fermion model
and then non-abelian bosonizing local operators. However, CFTs with the same
current algebras actually can have more than one IR solution of CFTs.  Their
distinction is clear in the operator content, but rather subtle in the
Lagrangians.\cite{francesco2012conformal} Here, we take a new route by directly
constructing partition functions of the critical theory based on symmetries,
anomalies and constraints on local operators. The complete physical data are
encapsulated in the partition function of a CFT. 

The partition function of a CFT defined on a Euclidean spacetime torus is given by
\begin{align}
Z(\tau) = \Tr e^{-\Im (\tau) H+\ii \Re (\tau) K}\label{Ztau}
\end{align}
where $H$ is the Hamiltonian, $K$ is the total momentum operator. $\tau$ is a
formal parameter describing the shape of the torus. $\Re (\tau)$ and $\Im
(\tau)$ can be viewed as the periodicity in space and time direction
respectively. All fields in $1+1$d CFT can be factorized into right- and
right-moving fields, $f(z=x^1-\ii x^0)$ and left-moving fields $\bar f (\bar
z=x^1+\ii x^0)$, where $x^1$ and $x^0$ are space and time coordinate. The
critical point under consideration is also  invariant under $G=SO(3)\times
U(1)$ onsite symmetry. In the low energy, $G$ can act on $f(z), \bar f(\bar z)$
independently. The symmetry is thus enlarged to $G\times \bar G$. All operators
in CFTs with this enlarged symmetries can be constructed from the currents and
representations of $su(2)_k$ and $u(1)_M$ current algebras, (which we explain
below, also see Appendix \ref{cft}) where the integer number $k$ and $M$ are
called levels. That levels are integers is a property of rational CFTs. IR fixed
points of WZW models with non-Abelian group are believed to be rational based
on one-loop renormalization group computation.\cite{witten1984non} Levels and
symmetry charges are restricted by $U(1)\times U_t(1)$ mixed anomaly, a
character of boson condensed state. 

 A necessary condition for a CFT to describe a bosonic critical theory of a lattice model is that the Euclidean partition function must be invariant, under modular transformation of the torus,
\begin{align}
\calS: Z(\tau)\rightarrow Z\left( -\frac{1}{\tau}\right),\quad \calT: Z(\tau)\rightarrow Z(\tau+1).
\end{align}
Namely tori with shape $\tau$, $-\frac{1}{\tau}$ and $\tau+1$ are all conformally equivalent and share identical (bosonic) partition functions.

Following the working assumption that CFTs with modular invariant partition functions always correspond to $1+1$d quantum (bosonic) models on a lattice, we construct UV completable CFTs. In particular, there is one set of CFTs, each has a single relevant symmetric direction. They describe critical points of lattice models with $U(1)\times SO(3)\times U_t(1)$ symmetry.

The simplest one has the following partition function
\begin{align}
\label{Z1tau}
&Z_A(\tau)= \left|\chi^{u1_4}_0\chi^{su2_2}_0+\chi^{u1_4}_2\chi^{su2_2}_1\right|^2\\
&+\left|\chi^{u1_4}_0\chi^{su2_2}_1+\chi^{u1_4}_2\chi^{su2_2}_0\right|^2+\left|\chi_1^{u1_4}+\chi_3^{u1_4}\right|^2\left|\chi_{\fh}^{su2_2}\right|^2
\nonumber 
\end{align}

\begin{table}[tb]
\centering
\begin{tabular}{  |c|c|c|c|c|}
\hline
operators & spin & charge & $U_t(1)$ & $h,\bar h$\\
\hline 
\rule{0pt}{3.4ex}$ \ee^{\pm \ii \vphi} V^{su2_2}_{1,m} $ & $1$ & $\pm 2$ & $\pm \frac12 k_B $ & $1,0$ \\
\hline 
\rule{0pt}{3.4ex}$ \ee^{\pm \ii \bar \vphi} \bar V^{su2_2}_{1,m} $ & $1$ & $\pm 2$ & $\mp \frac12 k_B$ & $0,1$ \\
\hline 
\rule{0pt}{3.4ex}$\ee^{\pm \ii (\vphi+ \bar\vphi)} $ & $0$ & $\pm 4$ & 0 & $\frac12,\frac12$ \\
\hline 
\rule{0pt}{3.4ex}$\ee^{\pm \ii (\vphi- \bar\vphi)} $ & $0$ & $0$ & $\pm k_B$& $\frac12,\frac12$  \\
\hline 
\rule{0pt}{3.4ex}$\ee^{\pm \ii \vphi} \bar V^{su2_2}_{1,m} $ & $1$ & $\pm 2$ & $\pm \frac12 k_B$& $\frac12,\frac12$  \\
\hline
\rule{0pt}{3.4ex}$ V^{su2_2}_{1,m} \ee^{\pm \ii \bar\vphi} $ & 1& $\pm 2$ & $\mp \frac12 k_B$ & $\frac12,\frac12$  \\
\hline
\rule{0pt}{3.4ex}$ V^{su2_2}_{1,m}  \bar V^{su2_2}_{1,m'} $ & 0, 1, 2 & 0 & $0$& $\frac12,\frac12$  \\
\hline
\rule{0pt}{3.4ex}$\ee^{\pm \ii \frac{\vphi+\bar\vphi}{2}}V^{su2_2}_{\frac12,\pm \frac12} 
\bar V^{su2_2}_{\frac12,\pm \frac12}$
 & $0,1$ & $\pm 2$ & 0 & $\frac5{16},\frac5{16}$  \\
\hline
\rule{0pt}{3.4ex}$\ee^{\pm \ii \frac{\vphi-\bar\vphi}{2}}V^{su2_2}_{\frac12,\pm \frac12} 
\bar V^{su2_2}_{\frac12,\pm \frac12}$
 & $0,1$ & $0$ & $\pm \frac12 k_B$& $\frac5{16},\frac5{16}$  \\
\hline
\end{tabular}
\caption{
The local operators (the primary fields) in the $Z_A$-CFT, which is a version
of the $su(2)_2\oplus u(1)_4 \oplus \bar{su(2)_2}\oplus \bar{u(1)}_4$ CFT. And their spin, $U(1)$ charge, and $U_t(1)$ charge.}
\label{za}
\end{table}

Here, the CFT is generated by $su(2)_2\oplus u(1)_4\oplus \bar{su(2)}_2\oplus \bar{u(1)}_4$ current algebra acting on right and left movers. In general for positive integral level $k$ and $M$, the partition function $Z(\tau)$ of a rational (finite) CFT can be organized into a finite sum, $Z(\tau)=\sum_{\mu,\nu}M_{\mu\nu}\chi_\mu\bar \chi_\nu$. It means physical excitations are (finite number of) representations of both right and left current algebras, labeled by $\mu,\nu$ respectively. $\chi_\mu$ ($\bar \chi_\nu$) is a so-called character associated with the primary field of right(left) current algebras. The character $\chi_\mu$ encodes the spectrum of excitations that are also the representation $\mu$ of current algebras.  The multiplicity $M_{\mu\nu}$ must be a non-negative integer, representing the number of times the excitations $(\mu,\nu)$ appear in the spectrum. 

The $u(1)_M$ and $su(2)_k$ CFTs are well-established.\cite{francesco2012conformal} Concisely, the character $\chi_m^{u1_M}$ of $u(1)_M$ CFT, where $0\leq m<M$ and $R^2=M$, (see supplementary material for the exact form) corresponds to primary fields of $u(1)_M$ current algebra, 
\begin{align}
\label{u1Op}
\ee^{\ii (\frac{m}{R}+n R) \varphi (z)}.
\end{align} The $u(1)_M$ primary fields represent the gapless charge excitations and carry the $U(1)$ charge quantum number. In addition, they also carry the $U_t(1)$ quantum number (or large momentum, explained later). The character $\chi_j^{su2_k}$ of $su(2)_k$ CFT, where $j=0,\frac{1}{2},\cdots, \frac{k}{2}$, corresponds to the primary fields of $su(2)_k$ CFT, denoted as $V_{j,m}^{su2_k}$, where $m=-j,-j+1,\cdots, j$. They represent gapless spin excitations, carrying spin quantum number. 

$Z_A$ is modular invariant under $\calS$ and $\calT$ transformation  shown in
Appendix \ref{characters}. Now we show that $Z_A$-CFT in (\ref{Z1tau})
describes the critical point between the two superfluids. The critical theory
between the two superfluids is first a spin-1 charge-2 boson condensing state.
All gapless excitations must carry integer spins and even charges. The minimal
charge is 2 and the minimal spin is 1. Generically, CFTs constructed from
$su(2)_k\oplus u(1)_M \oplus \bar{su(2)}_k\oplus \bar{u(1)}_M$ current algebra
for a fixed level $k$ and $M$ are not unique, but can have many versions
corresponding to different modular invariant partition functions. They are all
candidates to describe charge-$2$ spin-$1$ boson condensing states. These
states may even be transmuted into one and other under interactions (that are
marginal and symmetric). However, some condensing states are more stable than
others. First, the integral spin sector $su(2)_k$ with smallest possible $k$ is
most stable. Since $su(2)_k$ CFT with smaller $k$ has a smaller central charge
$\frac{3k}{k+2}$ and is more stable under RG flow where the central charge can
monotonically decrease or stay the same at best. Second, the spin sector is
most naturally realized by $su(2)_k$ CFT with even $k$. In fact, if we gap the
charge sector by ($U(1)$ or $U_t(1)$ symmetry breaking) perturbations, the
charge sector is insulating, the spins effectively become a spin-$1$ chain.
On a spin-integer chain, the phase transition respecting $SO(3)$ and
translation symmetry can only be $su(2)_k$ CFTs with even $k$.
\cite{tsvelik1990tsvelik, furuya2017symmetry}\footnote{The evenness or oddness
of the level in $su(2)_k$ theories is a global property, that can be detected
by coupling to background $\ZZ_2$ gauge field. Apply a $\ZZ_2\subset SU(2)$
symmetry twist to a consistent theory to describe the critical point with
$SO(3)$ symmetry produces another anomalous free theory. However, only the
$su(2)_k$ CFT with even $k$ satisfies this consistent condition and gives a
modular invariant CFT\cite{francesco2012conformal}. $su(2)_k$ CFTs with odd $k$
suffer from a global $\ZZ_2$
anomaly.\cite{gepner1986string,numasawa2018mixed,Lin:2019kpn} For example,
$su(2)_1$ CFT under a $\ZZ_2$ symmetry twist gives rise to purely left (or
right) moving spin-$\frac{1}{2}$ excitations $V_{\fh}^{su2_1}$ which cannot
exist in a spin-1 bosonic system. This global $\ZZ_2$ anomaly is a property
sustained under renormalization flow\cite{furuya2017symmetry}.}We conclude that
$su(2)_2$ CFT for the spin sector is the most stable. Its primary fields can
carry spin $0,\frac{1}{2}$ and $1$.

The charge sector is built from $u(1)_M$ CFTs. Charge excitations are described
by $u(1)_M$ primary field (\ref{u1Op}). The purely left charge-2 spin-1 boson
$\ee^{\ii a\phi}V_1^{su2_2}$ must have conformal spin $h-\bar h=1$ and that
fixes $\ee^{\ii a\varphi}$ to have scaling dimension $h=\fh$ and $a=1$. (In
this paper, we adopt the normalization such that $\langle\varphi (z)\varphi
(z')\rangle\sim (z-z')^{-1}$ and similarly for left-moving field $\bar
\varphi$.) The low energy field can always factorized into left and right
movers, we allow charge-1 primary field in left (right) $u(1)_M$ current
algebra, which would be $\ee^{\ii \frac{1}{2}\phi}$, therefore the minimal
level we need is $M=4$ by comparing with (\ref{u1Op}). (Other consisitent
levels are shown in Appendix \ref{z2orbifold}.) $\ee^{\ii \frac{1}{2}\varphi}$
and $\ee^{\ii \frac{1}{2}\bar \varphi}$, each carries charge $q=1$ so that
there are local excitations $\ee^{\ii
\frac{\varphi+\bar\varphi}{2}}V_{\frac{1}{2},l}^{su2_2}\bar
V_{\frac{1}{2},l}^{su2_2}$ carrying the minimal charge $2$ and spin $0$ or $1$.

The critical point is also invariant under $U_t(1)$ translation symmetry. The
$U_t(1)$ quantum number can only be carried by $u(1)_M$ primary fields, to be
consistent with fusion rules among primary fields. Boson condensed states have
a mixed $U(1)$ and $U_t(1)$ anomaly. This property is useful to restrict the
levels of $u(1)_M$ CFTs and fix the $U_t(1)$ charge in each critical theories.
We first define the $U_t(1)$ ``charge''. In the gapless bosonic system in 1d,
there are low energy excitations carrying small momentum $\sim 0$, representing
for example phonon modes. There are also low energy excitations with large
momentum.  They are center of mass motions of all bosons that can be generated
by Galilean boost. Each boson is excited with a momentum kick of the order
$\frac{2\pi}{L}$, and the total momentum change is of the order
$N\frac{2\pi}{L}\equiv k_B=\frac{2\pi}{a}\nu$, proportional to the boson
density, where $N$ is total boson number, $a$ is lattice spacing and
$\nu=\frac{N}{N_L}$ is the boson filling number. We denote momentum of the
center of mass motion as the $U_t(1)$ ``charge'' $q_t$, which is always of
order $k_B$. Excitations with momentum $\ll k_B$ has $q_t=0$. The center of
mass motion can also be viewed as an excitation that changes the boundary
condition of the field in $u(1)_M$ CFT from a periodic one to
$\phi(L)=\phi(0)+\frac{L}{N}	q_t$, which is sometimes called a large gauge
transformation. (More physical discussion is in Appendix \ref{anomaly}.) 

\begin{figure}[tb]
\begin{center}
\includegraphics[width=2.5in]{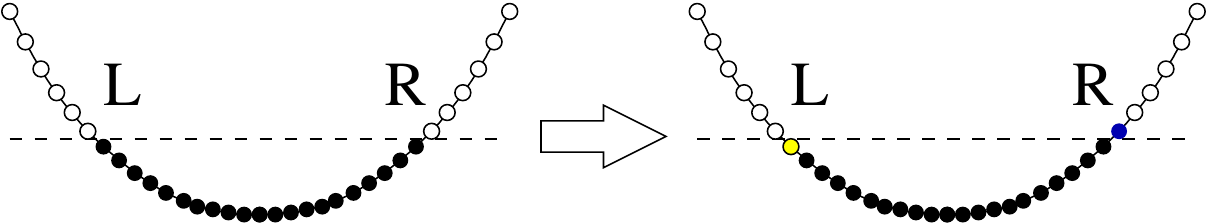} 
\end{center}
\caption{
\label{boost}
For right-moving fermions with charge $q$ and $U_t(1)$-charge $q_t$ on a ring,
adding a $2\pi/q$-flux of $U(1)$ through the ring will creat a fermion and add a $U_t(1)$-charge $q_t$.  So the $U(1)\times U_t(1)$ mixed anomaly for
the right-moving fermions is given by $q q_t$, \ie $2\pi$-flux of $U(1)$ will
create a $U_t(1)$-charge $qq_t$.  Similarly, for left-moving fermions with
charge $q$ and $U_t(1)$-charge $-q_t$, the $U(1)\times U_t(1)$ mixed anomaly is
$q q_t$.  }
\end{figure}

A boson condensed state has a characteristic feature. Modes with the  center of
mass momentum can be excited when pumping $U(1)$ flux through the ring. To
measure this mixed anomaly in charge-2 spin-1 boson condensing state, we turn
on a $U(1)$-flux $\Phi=2\pi $ through the ring. The process boosts each boson
that carries $U(1)$ charge $q=2$  by a momentum
$\frac{q\Phi}{L}=\frac{4\pi}{L}$, thus in total changes $q_t$ by $2k_B$.   Low
energy theories shall carry this amount of anomaly.  The $U_t(1)$ charge of
local operators is determined in accord with $U(1)$ charge via matching the
mixed anomaly. The anomaly of the operator $\ee^{\pm \ii (\varphi-\bar
\varphi)}$, carrying total charge $q=0$ and non-trivial $q_t$, is most manifest
to compute. The right-moving part $\ee^{\ii \varphi}$ has scaling dimension
$(h,\bar h)=(\frac{1}{2},0)$ and thus represents a chiral fermion. Its left
charges are $q=2$ and $q_t$ to be determined. The right-moving part
$\ee^{\ii\varphi}$ has $U(1)$ charge $q$ and $q_t$ (and similarly, the
left-moving part $\ee^{-\ii\bar \varphi}$ has $-q$ and $q_t$ charges) . Adding
$2\pi$-flux of $U(1)$ would create $q$ right-moving fermions and change
$U_t(1)$ charge by $qq_t$; and annihilate $q$ left-moving fermions and change
$U_t(1)$ charge by $qq_t$, as illustrated in Fig.\ref{boost}. The total change
of $U_t(1)$ charge is $2qq_t$. That it equals $2k_B$ fixes
$q_t=\frac{1}{2}k_B$. It follows that in total, the operator $O_{e,b}(z,\bar
z)=\ee^{\ii \left(\frac{e}{\sqrt{M}}+\frac{b}{2}	\sqrt{M}\right)\varphi+
\ii \left(\frac{e}{\sqrt{M}}-\frac{b}{2}	\sqrt{M}\right)\bar \varphi}$
has charges
\begin{align}
q=2e, \quad q_t= b k_B
\label{charges}
\end{align}

The partition function $Z_A$ determines all allowed local excitations, operators corresponding to primary fields, organized in Table. \ref{za}. The local operators carry spin, charge, and translation $U_t(1)$ charge. 
The spin quantum number is the same as the spin $j$ of primary fields in $su(2)_k$ CFT. 
All local operators in Table \ref{za} carry even charges. However, there are excitations with fractionalized $U_t(1)$ charge $\pm \frac{1}{2}k_B$. Furthermore, the excitation carries spin-0, and charge-0 (left charge $\pm 2$ and right charge $\mp 2$. Such an excitation is indeed created by adding $\pi$-flux of $U(1)$ through the ring. 

The CFT with $Z_A$ is a representative of a series of $su(2)_2\oplus u(1)_{M}\oplus \bar{su(2)}_2\oplus \bar{u(1)}_M$ CFTs with $M=0 \mod 4$, which are all $Z_2$ orbifold CFTs. We list the operator content in Table \ref{M4} and explained in Appendix \ref{z2orbifold}. 
We see that minimal charge is $\pm 2$, minimal $q_t$ charge is $\pm\frac{2}{M}$ and the mixed anomaly is $2k_B$. In each theory, there is only one relevant operator $\calO$ symmetric under $SO(3)\times U(1)\times U_t(1)$ symmetry.

As shown in Table \ref{za}, the single symmetric relevant operator is
\begin{align}
\calO=\sum_{l=x,y,z}V_{1,l}^{su2_2}\bar V_{1,l}^{su2_2}
\end{align} 
that carries trivial quantum numbers and total scaling dimension $\Delta=h+\bar h=1<2$. We would like to show that this perturbation drives the critical state to a superfluid state where all spin excitations are gapped. 

\noindent\textbf{Topological continuous phase transition between  superfluids} We consider the two phases connected by the critical theory $Z_A$ perturbed by the relevant operator $\calO$. First the perturbation is all within the spin sector $su(2)_2\oplus \bar{su(2)}_2$ with central charge $c+\bar c=3$. The spin sector can be described by three Majorana fermions that form the spin-1 representation of $SU(2)$. The relevant operator $\calO$ corresponds to $SU(2)$ symmetric mass term of Majorana fermions with scaling dimension $(\frac{1}{2},\frac{1}{2})$. Adding the mass term can gap out all spin excitations. Therefore, the critical state described by $su(2)_2\oplus u(1)_4\oplus \bar{su(2)}_2\oplus \bar{u(1)}_4$ , a spin-1 charge-2 boson condensed state, is unstable against the gap openning for spin excitations. 

At incommensurate filling, translation symmetry cannot be spontaneously broken.
The gapped spin sector is therefore $SO(3)\times U_t(1)$ symmetric. Based on
$1+1d$ SPT classification, the charge-2 spin-1 bosons can form two stable phases
where charge operators have algebraic correlations, and all spin operators have
exponential decaying correlations. One phase is the TSF and the other is the
nTSF\cite{essler2009spin, zhou2003spin} introduced in the introduction.



The critical theory is also supported by its analog on the antiferromagnetic
spin-1 chain. There, the critical point between the AKLT phase and the
dimerized phases that spontaneously break translation symmetry are believed to
be described by $su(2)_2\oplus \bar{su(2)}_2$ CFT, and the transition is driven
by $\calO$ operator.\cite{affleck1987critical,tsvelik1990tsvelik} This result
is confirmed by DMRG calculation\cite{michaud2012antiferromagnetic}. We believe
that in the superfluid case with all density configurations summed over, the
spin part of the transition is still described by $su(2)_2\oplus \bar{su(2)}_2$
CFT. Combining that the charge part is described by $u(1)_M\oplus\bar  u(1)_M$.
$Z_A$-CFT can be tested by DMRG computation in the model (\ref{ham}) at the
critical point. From the minimal exponent of boson to be $\frac{5}{8}$ to
demonstrate that this is a critical point described by a $u(1)\oplus su(2)_2$
CFT. One can find the  minimal exponent of spin-1 charge-0 with nearly
vanishing momentum to be $1$, demonstrating that between TSF and nTSF, the
critical theory is $u(1)_4\oplus su(2)_2$ CFT with partition function $Z_A$ (as
opposed to $Z_B$ and $Z_C$  described in Appendix \ref{diagonalZ}.)

Since superfluid phases are gapless phases, they can also be described by modular invariant CFTs. To obtain the partition functions of two superfluids, we start with the critical $Z_A$-CFT. We may assume that the pure $u(1)_4$  operators $\ee^{\pm \ii 2\varphi}$ that are mutually local with the gapping operator in the spin sector remain. They appear in the CFTs that describe the two superfluid phases and carry same quantum numbers. This motivates us to use $u(1)_4$ CFT to describe two superfluids. There are two modular invariant partition functions for $u(1)_4$ CFT's. We believe they describe the two superfluid phases. The partition function describing the topological superfluid phase is 
\begin{align}
Z_{\text{top}}^{\text{SF}}(\tau)=\sum_{i=0}^3|\chi_i^{u1_4}(\tau)|^2
\end{align}
 \begin{table}[tb]
 \centering
 \def\arraystretch{1.7}
 \begin{tabular}{|c | c | c | c | c|}
 \hline
 operators & spin & charge & $U_t(1)$ & $h,\bar h$ \\
 \hline
 $\ee^{\pm \ii m \frac{\varphi+\bar \varphi}{2}}$ & $0$ & $\pm 2m$ & $0$ & $\frac{m^2}{8},\frac{m^2}{8}$ \\
 \hline
  $\ee^{\pm \ii 2 \varphi}$ & $0$ & $\pm 4$ & $\pm k_B$ & $2,0$ \\
   \hline
  $\ee^{\pm \ii 2 \bar \varphi}$ & $0$ & $\pm 4$ & $\mp k_B$ & $0,2$ \\
  \hline
 \end{tabular}
  \caption{The local operators in the $Z_\text{top}^\text{SF}$-CFT. }
 \label{SFtop}
 \end{table}
 \begin{table}[tb]
 \centering
 \def\arraystretch{1.7}
 \begin{tabular}{|c | c | c | c | c|}
 \hline
 operators & spin & charge & $U_t(1)$ & $h,\bar h$ \\
 \hline
 $\ee^{\pm \ii m \frac{\varphi-\bar \varphi}{2}}$ & $0$ & $0$ & $\pm \frac12 k_B $ & $\frac{m^2}{8},\frac{m^2}{8}$ \\
 \hline
  $\ee^{\pm \ii 2 \varphi}$ & $0$ & $\pm 4$ & $\pm k_B$ & $2,0$ \\
   \hline
  $\ee^{\pm \ii 2 \bar \varphi}$ & $0$ & $\pm 4$ & $\mp k_B$ & $0,2$ \\
  \hline
 \end{tabular}
  \caption{The local operators in the $Z_\text{tri}^\text{SF}$-CFT. }
 \label{SFtri}
 \end{table}
Its operator content is given in Table \ref{SFtop}. The $U(1)\times U_t(1)$ mixed anomaly remains $2k_B$. The minimal charge is $2$, as carried by the operator $\ee^{\ii\frac{1}{2}(\varphi+\bar \varphi)}$. We stress that while the low energy theory of TSF can be described by $u(1)_4$ CFT, the phase itself requires the gapped spin-1 sector to reveal its nature of fractionalization. The non-topological superfluid is described by modular invariant partition function 
\begin{align}
Z_\text{tri}^\text{SF}= 
 |\chi^{u1_4}_0|^2 +\chi^{u1_4}_1 \bar \chi^{u1_4}_3
+|\chi^{u1_4}_2|^2 +\chi^{u1_4}_3 \bar \chi^{u1_4}_1.
\end{align}
Its operator content is given in Table \ref{SFtri}. The minimal charge is $4$, as carried by the operator $\ee^{\ii(\varphi+\bar \varphi)}$. 

We have shown that $Z_A$-CFT is consistent to describe the critical point
of charge-2 spin-1 bosons with antiferromagnetic interaction. Besides the proposal of a boson t-J model, we also provide a construction of the $Z_A$-CFT
from a spin-$\fh$ fermion system in Appendix \ref{ferm}. We show that all fermionic excitations are gapped, the system is effectively spin-1 bosonic and described by $Z_A$-CFT.


Wenjie Ji thanks Zhen Bi, Liujun Zou, Max Metlitski, Ethan Lake, Dominic Else for fruitful discussions. This research is partially supported by NSF grant DMR-1506475 and DMS-1664412.

\vfill
\break

\appendix

\section{Modular invariant partition functions}\label{cft}
In the low energy theory of a critical point with onsite symmetry $G$ and emergent conformal symmetry, $1+1$d conformal symmetry ensures all fields can be factorized into left-moving and right-moving fields, including the conserved currents associated with $G$. Left moving and right moving currents are conserved separately with scaling dimensions $1$ and vanishing anomalous dimensions. These currents generate left and right current algebras. 

\subsection{Current algebras}
The currents $J(z)= J^a (z) T_a, \bar J(\bar z)=\bar J_a (\bar z)T_a$, where $T_a$ are generators of the Lie group $G$ or the corresponding Lie algebra $g$ are conserved, 
\begin{align}
\partial_{\bar z} J(z)=0, \quad \partial_z \bar J (\bar z)=0
\end{align}
The generators of a current algebra $g_k$ at level $k$ can be defined via the operator product expansion of current fields, 
\begin{align}
J^a (z) J^b (w)\sim \frac{k\delta^{ab}}{(z-w)^2}+\frac{{f^{ab}}_c J^c(w)}{z-w}
\end{align}
From this, one see that the generators are primary fields with scaling dimension $(h, \bar h)=(1,0)$ and are also called currents. The primary fields $\mu$ are representations $\mu$ of the current, 
\begin{align}
J^a(z) \phi_{[\mu],i}(w)\sim \frac{t^a_{[\mu],ij}\phi_{[\mu],j} (w)}{(z-w)}
\end{align}
where $t^a_{[\mu]}$ is the matrix of $J^a$ in the representation $\mu$. Though operator product expansion, the currents and primary fields generate all operators in the current algebra (and all possible corresponding CFTs).\cite{francesco2012conformal} 


\subsection{Characters of current algebras}
\label{characters}
$u(1)_M$ current algebra is generated by the current $\partial_z \varphi (z)$ and $\ee^{\ii \sqrt{M}\varphi}$. The primary fields  of the current algebra are $\ee^{\ii \frac{m}{\sqrt{M}}\varphi}, 0\leq m\leq M-1$. The character $\chi^{u1_M}_m$ of $u(1)_M$ CFT is given by 
\begin{align}
\label{u1chi}
 \chi^{u1_M}_m(\tau) =& \eta^{-1}(q)
\sum_{n=-\infty}^\infty q^{\frac{1}{2} (\frac{m}{R}+n R)^2} ,
\end{align}
where $\eta(q) = q^{\frac{1}{24}} \prod_{n=1}^\infty (1-q^n)$ is the Dedekind function, $0 \leq m <
M$, and $R^2=M$. Under modular transformation $\calS$ and $\calT$, the characters transform as follows,
\begin{align}
\begin{split}
 &\chi^{u1_M}_i(-\frac{1}{\tau}) =
\sum_j \calS_{ij} \chi^{u1_M}_j(\tau),\ \ 
\calS_{ij}=\frac{1}{\sqrt{M}} \ee^{-\ii 2\pi \frac{ij}{M} },\\
 &\chi^{u1_M}_i(\tau+1)= \ee^{-\ii \frac{2\pi}{24}}\ee^{\ii 2\pi  \frac{i^2}{2M}}\chi^{u1_M}_i(\tau)
\end{split}
\end{align} 
$su(2)_k$ is generated by $J(z)=J^a(z)T_a$, where $T_a, a=1,2,3$ are generators of $su(2)$ Lie algebra. The character $\chi_{j}^{su2_k}(\tau)$ of $su(2)_k$ CFT is given by
\begin{align}
\label{su2chi}
\chi_{j}^{su2_k}(q)=&\frac{q^{(2j+1)^2/4(k+2)}}{[\eta (q)]^3}
\\
&\cdot \sum_{n\in\Z} \left[2j+1+2n(k+2)\right] q^{n[2j+1+(k+2)n]}
\nonumber 
\end{align}
where $j\in \left\{0,\frac{1}{2},\cdots,\frac{k}{2}\right\}$. Their transformations under $\calS$ and $\calT$ are
\begin{align}
&\chi^{su2_k}_j(-1/\tau)=\sum_{l\in\cP}\calS_{jl}\chi^{su2_k}_l(\tau),\nn\\
&\calS_{jl}=\sqrt{\frac{2}{k+2}}\sin \left[ \frac{\pi (2j+1)(2l+1)}{k+2}\right]\\
&\chi^{su2_k}_j(\tau+1)=\ee^{-\ii 2 \pi\frac{3k}{24(k+2)}}\ee^{\ii 2\pi \frac{j(j+1)}{k+2}}\chi^{su2_k}_j(\tau). \nn
\end{align}

Based on the above $\calS$ and $\calT$ matrices, one can check that $Z_A,Z_B,Z_C$ and all orbifold models in Appendix \ref{z2orbifold} are modular invariant. 

\subsection{Other modular invariant solutions}\label{diagonalZ}
There are, in addition, two types of $su(2)_2\oplus u(1)_M\oplus \bar{su(2)}_2\oplus \bar{u(1)}_M$ CFTs that are modular invariant. Their partition functions are the products of modular invariant partition functions of the $u(1)_M$ and the $su(2)_k$ part, 

\begin{align}
\begin{split}
Z_B=& \Big( \sum_{i=0}^{M-1} |\chi^{u1_M}_i|^2 \Big) \Big( \sum_{j=0,\frac{1}{2},1} |\chi^{su2_2}_j|^2 \Big),
\\
Z_C=& \Big( \sum_{i=0}^{M-1} \chi^{u1_M}_i \bar \chi^{u1_M}_{M-i} \Big)
\Big( \sum_{j=0,\frac{1}{2},1} |\chi^{su2_2}_j|^2 \Big)
\end{split}\label{zb}
\end{align}

$Z_B$ is the most familiar partition function for WZW model with $U(1)$ at level $M$ and $SU(2)$ at level $2$. This type of partition function is called diagonal, since the operators carry the same left and right representations of current algebras. 
\begin{table}[tb]
 \centering
 \def\arraystretch{1.7}
 \begin{tabular}{|c | c | c | c | c|}
 \hline
 operators & spin & charge & $U_t(1)$ & $h,\bar h$ \\
 \hline
 $\ee^{\pm \ii m \frac{\varphi+\bar \varphi}{\sqrt M}}$ & $0$ & $\pm 2m$ & $0$ & $\frac{m^2}{2 M},\frac{m^2}{2M}$ \\
 \hline
  $\ee^{\pm \ii \sqrt M \varphi}$ & $0$ & $\pm M$ & $\pm k_B$ & $\frac{M}{2},0$ \\
   \hline
  $\ee^{\pm \ii \sqrt M \bar \varphi}$ & $0$ & $\pm M$ & $\mp k_B$ & $0,\frac{M}{2}$ \\
  \hline
  \hline
  $V_{\fh}^{su2_2}\bar V_{\fh}^{su2_2}$ & $0,1$ & $0$ & $0$ & $\frac{3}{16},\frac{3}{16}$ \\
    \hline
  $V_{1}^{su2_2}\bar V_{1}^{su2_2}$ & $0,1,2$ & $0$ & $0$ & $\frac{1}{2},\frac{1}{2}$ \\
  \hline
 \end{tabular}
  \caption{The local operators in the $Z_B$-CFT, where $1\leq m<M$.}
 \label{zb}
 \end{table}
 \begin{table}[tb]
 \centering
 \def\arraystretch{1.7}
 \begin{tabular}{|c | c | c | c | c|}
 \hline
 operators & spin & charge & $U_t(1)$ & $h,\bar h$ \\
 \hline
 $\ee^{\pm \ii m \frac{\varphi-\bar \varphi}{\sqrt M} }$ & $0$ & $0$ & $\pm mk_B$ & $\frac{m^2}{2M},\frac{m^2}{2M}$ \\
 \hline
  $\ee^{\pm \ii \sqrt M \varphi}$ & $0$ & $\pm 2$ & $\pm \frac{M}{2}k_B$ & $\frac{M}{2},0$ \\
   \hline
  $\ee^{\pm \ii \sqrt M \bar \varphi}$ & $0$ & $\pm 2$ & $\mp \frac{M}{2}k_B$  & $0,\frac{M}{2}$ \\
  \hline
  \hline
  $V_{\fh}^{su2_2}\bar V_{\fh}^{su2_2}$ & $0,1$ & $0$ & $0$ & $\frac{3}{16},\frac{3}{16}$ \\
    \hline
  $V_{1}^{su2_2}\bar V_{1}^{su2_2}$ & $0,1,2$ & $0$ & $0$ & $\frac{1}{2},\frac{1}{2}$ \\
  \hline
 \end{tabular}
  \caption{The local operators in the $Z_C$-CFT, where $1\leq m<M$.}
 \label{zc}
 \end{table}

Their operator contents in the charge and spin sectors, which are totally decoupled are given in Table
\ref{zb} and \ref{zc}. Generic operators are products of operators in charge and spin sectors. Quantum numbers are determined similarly as in Table \ref{za}, based on four requirements: (a) the minimal
$U(1)$-charge is 2 and other $U(1)$ charges are multiple of 2; (b) the mixed
$U(1)\times U_t(1)$ anomaly is $2k_B$; (c) there exists an operator with
spin-0, charge-0, and $U_t(1)$-charge $k_B$; (d) there exists an operator with
spin-1, charge-2, and $U_t(1)$-charge $0$.  The above conditions can be
satisfied only if $M$ is even.


$Z_B$-,$Z_C$-CFTs also preserve $U(1)\times SO(3)$ symmetry and may lie within the phase boundary between the topological and non-topological superfluid. However, both theories have two symmetric relevant operators, singlets composed from $V_{\fh,l}\bar V_{\fh,l}$ and $V_{1,m}\bar V_{1,m}$. They are, therefore, at best multi-critical points between the two superfluids. 

\section{Charges in superfluids and $U(1)\times U_t(1)$ mixed anomaly}\label{anomaly}

Consider a superfluid in 1+1d (one could view it as the limit of a superfluid on a solid torus whose radius of the cross section going to zero). The superfluid described by $\phi(x)$ field can carry different center of mass velocity $v_s=\partial_x \phi(t,x)$. The momentum is $P=Nv_s$, (where we set the boson mass to 1.) and the superfluid field has boundary condition $ \phi(L)=\phi (0)+v_sL=\phi(0)+\frac{P}{\rho}, \rho=\frac{N}{L}$. One cannot untwist the superfluid to $\phi(L)=\phi(0)$ with $v_s=0$, without tearing the superfluid. The discontinuity of the compact bosonic field must be $v_sL=2\pi R m$, where $m$ is an integer, and $R$ may be different from $1$ describes the interaction strength in the superfluid. Thus, $P=k_B Rn$, where we denote $k_B=\frac{2\pi N}{L}$. 

One can use $u(1)_M$ CFT to describe the superfluid when $M=R^2$ is an integer.  The Hilbert space decompose into sectors accordingly, 
\begin{align}
\calH=\oplus_{\mu\in\ZZ} \calH_\mu, \quad \calH_\mu: \phi(L)=\phi(0)+2\pi R\mu
\end{align}
 Exications in each sector are operators $\ee^{\ii \frac{e}{R}\phi (z,\bar z) }$ of charge $e\in\ZZ$, where we now use the left- and right-moving coordinates, $\phi(z,\bar z)= \varphi(z)+\bar \varphi (\bar z)$. 

Operators that maps states in one sector to the other is $O_{0,b}=\ee^{\ii \frac{b}{2}R(\varphi-\bar \varphi)}$. Consider an excitation $O_{e,0}=\ee^{\ii \frac{e}{R}\phi}$ with electric charge $e$ going around an operator $O_{0,b}$ will accumulate a phase of $\ee^{\ii 2\pi eb}$. That is $O_{e,0}(z+r\ee^{\ii 2\pi}, \bar z+r\ee^{-\ii 2\pi})O_b(z,\bar z)=\ee^{\ii 2\pi eb}O_{e,0}(z+r, \bar z+r)O_b(z,\bar z)$. This phase accumulation is the same as the electrical charged operator with changed boundary condition $O'_{e,0}(z,\bar z)=\ee^{\ii \frac{e}{R}\phi'(z,\bar z)}$ where $\phi'(L)=\phi'(0)+2\pi Rb$ comparing to $O_{e,0}=\ee^{\ii \frac{e}{R}\phi}$ where $\phi(L)=\phi (0)$. Since the radial direction of $z$ is the spatial direction for CFT in a complex plane. That is $\lim_{|\delta z|\rightarrow 0}O_{e,0}(z+\delta z, \bar z+\delta \bar z)O_{0,b}(z,\bar z)\equiv O'_{e,0}(z,\bar z)$. The magnetic charged operator describes the excitations that change the center of mass momentum in a superfluid. 

Creation of a vertex with vorticity $q_t$ can untwist the superfluid and carries away the momentum by an amount $q_t$.

Now we apply the $u(1)_M$ CFT to described the charge sector of the boson condensing state (on a ring) at the critical point in the paper. 
To determine the $U(1)$ charge $q$ and large momentum charge $q_t$ for excitation $O_{e,b}$. $O_{1,0}=\ee^{\ii \frac{1}{R}(\varphi+\bar \varphi)}$ has the minimal $U(1)$ charge, and we require it to be $2$. It determines $q=2e$. The large momentum charge $q_t$ is determined by matching $U(1)\times U_t(1)$ anomaly. Pumping $U(1)$ flux through the ring adds a large momentum $2k_B$. That is to change the boundary condition of $\phi (x)$ field by $\frac{2k_BL}{N}$, so the operator for pumping $2\pi$ flux is a magnetic operator with charge $m$ satisfying $\frac{2k_BL}{N}=2\pi R b$, therefore $b=\frac{2}{R}$,
\begin{align}
O_{0,\frac{2}{R}}=\ee^{\ii (\phi-\bar \phi)}
\end{align}
which has scaling dimension $\left(\frac{1}{2}, \frac{1}{2}\right)$. As described in the main text, the left part of $O_{0,\frac{2}{R}}$ has $e=\frac{\sqrt{M}}{2}$, $b=\frac{1}{\sqrt{M}}$, left charge $q=\sqrt{M}$, and left $q_t$. The total mixed anomaly is $2qq_t=2k_B$ determines $q_t=\frac{1}{\sqrt{M}}k_B$. Therefore, $q_t=b k_B$. We arrive at the charges formula (\ref{charges}) of all excitations, also listed in Table (\ref{M4}).

In the diagonal and orbifold $u(1)_M$ CFT, all vertex excitations are of the above form with $R=\sqrt{M}, e,b \in \ZZ$. In the charge conjugated $u(1)_M$ CFT, such as the charge sector in $Z_C$-CFT, all vertex excitations are of the above form with $R^*=\frac{2}{\sqrt{M}}, e,b\in \ZZ$. This ensures the minimal charge of local excitation remains to be $\pm 2$.

\section{Construction of the critical state of spin-1 boson
from a spin-$\fh$ fermion model}
\label{ferm}

Here we present a realization of the $Z_A$-CFT, starting from a free fermionic model, and then
turning on an interaction.  First, we will show that in an interacting phase, the
fermionic excitations are fully gapped and the low energy theory is composed of
purely spin-1 bosons formed by fermion pairs.  Next, we will show that the resulting spin-1 boson system is in a critical state described by $Z_A$-CFT. 
In this construction, it is manifest to compute the $U(1)$
and $U_t(1)$ quantum numbers of operators in $Z_A$-CFT. 

To begin with, the Hamiltonian of a spin-$\fh$
charge-1 electron model on 1D lattice is
\begin{align}
\label{Ham}
 H_0=\sum_i (-t c^\dag_{\al a,i+1} c_{\al a,i} +h.c.)
-\mu c^\dag_{\al a,i} c_{\al a,i}
\end{align}
where $\al=\up,\down$ is the spin index and $a=1,2$ is the flavor index. The
chemical potential $\mu$ is chosen such that the fermions are at incommensurate
filling. At low energy, the above model has two right-moving spin-$\fh$ fermions
with crystal momentum $k_F$ and two left-moving spin-$\fh$ fermions with crystal
momentum $-k_F$.  Note that $2k_F =2\pi\frac{n_F}{4}$ where $n_F$ is the total
density of the fermions. As the fermion pairs form the spin-1 bosons,
the boson density is $n_B=n_F/2$. Therefore, $k_B=4 k_F$.

The two right-moving spin-$\fh$ fermions are described by
$\psi_{\al a}$ where $\al$ and $a$ are the spin and flavor index as well.  Similarly, the two left-moving spin-$\fh$ fermions are described
by $\bar \psi_{\al a}$. The model is translational invariant and the role of crystal momenta here is analogous to $U(1)$
charges that are conserved mod $2\pi$.  The low energy effective Hamiltonian
density is given by
\begin{align}
\label{H0}
 \cH_0 = 
\psi_{\al a}^\dag(x) \ii v_0 \prt_x \psi_{\al a}(x) -
\bar \psi_{\al a}^\dag(x) \ii v_0 \prt_x \bar \psi_{\al a}(x) 
\end{align}
At low energy, generically the spin, flavor and charge degrees of freedom may
have different velocities, the model has an emergent $[SU(2)_s\times SU(2)_f
\times U(1)]_R \times [SU(2)_s\times SU(2)_f \times U(1)]_L$ symmetry for right
and left movers. Correspondingly, the right movers of the above system are
described by a CFT with total central charge $c=4$: $su2^s_2\oplus
su2^f_2\oplus u1^c$, where the excitations in $su2^s_2$ carry $SU(2)_s$ spin
quantum numbers, the excitations in $su2^f_2$ carry $SU(2)_f$ flavor quantum
numbers, and the excitations in ${u1}$ carry the $U(1)$ charges.  Similarly,
the left movers of the above system are described by a CFT with central charge
$\bar c = 4$: $\bar{su2^s_2}\oplus \bar{su2^f_2}\oplus \bar{u1_4^c}$. Indeed, the $U(N_s\times N_f)$ Dirac fermion theory can be bosonized to the $SU(N_s)_{N_f}\times SU(N_f)_{N_c}\times U(1)_{N_sN_f}$ WZW theory. The levels for $su(N_s)$ and $su(N_f)$ are determined from the operator product expansion of currents. \cite{gonzales1985low, affleck1986realization, frishman1993bosonization} The level for $U(1)$ part is fixed by the condition that null states constructed from currents must be primary fields.

The local operators in the theory are powers of the fermion operators
$\psi_{\alpha a},\ \bar{\psi}_{\alpha a}$. The fermion operators can be represented in terms of the primary fields of the above CFTs\cite{francesco2012conformal}, 
\begin{align}
\label{fer1}
\begin{split}
\psi_{\alpha a} &\sim \ee^{\ii\frac{\varphi_c}{2}}   
\sigma_s\ee^{\pm \ii \frac{\phi_s}{2}}\sigma_f\ee^{\pm \ii \frac{\phi_f}{2}} 
=\ee^{\ii\frac{\varphi_c}{2}}V^{su2^s_2}_{\frac12,\pm \frac12}V^{su2^f_2}_{\frac12,\pm \frac12}
 \\
\bar\psi_{\alpha a} &\sim \ee^{\ii\frac{\bar\varphi_c}{2}}   
\sigma_s\ee^{\pm \ii \frac{\bar\phi_s}{2}}\sigma_f\ee^{\pm \ii \frac{\bar\phi_f}{2}} 
=\ee^{\ii\frac{\bar\varphi_c}{2}}\bar V^{su2^s_2}_{\frac12,\pm \frac12}\bar V^{su2^f_2}_{\frac12,\pm \frac12}
\end{split}
\end{align}
Here, for right movers, which are functions of $z=\tau+\ii x$ ($\tau$ and $x$ are imaginary time and space coordinate), \\
\indent (1) $\varphi_c$ is the bosonic field in 
$ {u1^c}$ CFT,\\
\indent (2) $\eta_s, \si_s$ and $\phi_s$ are the Ising CFT fields and 
the bosonic field to represent $su2^s_2$ primary fields, \\
\indent (3) $\eta_f, \si_f$ and $\phi_f$ are the Ising CFT fields and the bosonic field to represent $su2^f_2$ primary fields,\\
\indent (4) $V_{\frac{1}{2},\pm \frac{1}{2}}^{su2_2^s}$ is the primary field of $su2_2^s$ that is also a spin-$\frac{1}{2}$ doublet. And similarly for $V_{\frac{1}{2},\pm \frac{1}{2}}^{su2_2^f}$.\\
The fields for the left movers, which are functions of $\bar z=\tau-\ii x$,
are analogously defined. (In this paper, all the bosonic fields are normalized
such that $\langle \phi (z_1)\phi (z_2)\rangle =-\ln (z_1-z_2)$, \ie the
scaling dimension of $\ee^{\ii \phi}$ is $\frac{1}{2}$.) As expected, the
fermion operators are charge-$1$ spin doublet, as well as flavor doublet. Their
scaling dimensions are $\frac{1}{2}$, which is in fact, what we use to determine the
$u1$ part of (\ref{fer1}). 

Next, we obtain the desired critical state by simply gapping the flavor sector
$su2^f_2 \oplus \bar{su2^f_2}$.  This can be achieved dynamically by adding
a strong repulsive interaction for the flavor charges, for example, 

\begin{align}
V_{1,l}^{su2_2^f}\bar V_{1,l'}^{su2_2^f}\sim &\psi_{\alpha,2}^\dagger \psi_{\alpha,1}\bar{\psi}_{\beta,1}^\dagger\bar{\psi}_{\beta,2}+\psi_{\alpha,1}^\dagger \psi_{\alpha,2}\bar{\psi}_{\beta,2}^\dagger\bar{\psi}_{\beta,1}\nn\\
&+\psi_{\alpha,i}^\dagger \sigma^3_{ij}\psi_{\alpha,j}\bar \psi_{\beta,i}^\dagger \sigma^3_{ij}\bar \psi_{\beta,j},
\label{fperb}
\end{align}
which is symmetric flavor-pairing and contains Luther-Emery pairing terms. More specifically, the term is a singlet under both $SU_s(2)$ and $\bar{SU_s(2)}$, symmetric under $U(1)$ and $\bar{U(1)}$, and a singlet under the diagonal subgroup of $SU_f(2)\times \bar{SU_f(2)}$, that is formed from the triplet under $SU_f(2)$ and $\bar{SU_f(2)}$. 

It can gap out all
the flavor fluctuations.  The resulting critical state is what we want, described by CFT
\begin{align}
su2^s_2\oplus u1 \oplus \bar{su2^s_2}\oplus \bar{u1}. 
\label{su2su1}
\end{align}
The total central charge is
$c=\frac32+1$ for right movers and $\bar c=\frac32+1$ for left movers.

Note that in the theory, all fermionic operators, shown in (\ref{fer1}),
carry half-integer flavor of $SU(2)_f$, while the low energy CFT
(\ref{su2su1}) contains only flavor-singlet excitations. It means all low
energy excitations are bosonic, and must carry even charges and integer spins. The low energy theory is, therefore, a \emph{bosonic} theory.
Furthermore, the critical state should be viewed as a state of charge-2 spin-1
bosons (\ie electron pairs).  It is interesting that such a ``superconducting''
state can be induced by a repulsive interaction in the flavor density channel.  

We like to argue that the above critical state, $su2^s_2\oplus u1 \oplus
\bar{su2^s_2}\oplus \bar{u1}$, actually corresponds to the spin-1 boson
condensed state for anti-ferromagnetic spin interaction.  In the spin-1 boson
condensed state, both charge $U(1)$ symmetry and spin $SO(3)$ symmetry tend to
be spontaneously broken.  But in 1+1D, a continuous symmetry cannot really be
broken.  Here we assume the spins of the bosons form an algebraic
anti-ferromagnetic order, \ie the neighboring bosons have opposite spins.  (For
ferromagnetic interaction, the spins can have a true long-range order that
spontaneously breaks the spin rotation symmetry.)  Thus the spin of the bosons
can develop only an algebraic long-range order.  We believe that the gapless
spin excitations in such a state are described by the conserved $su2$ spin
currents.  And the gapless charge excitations are described by the conserved
$u1$ charge currents.  Thus the critical state $su2^s_2\oplus u1 \oplus
\bar{su2^s_2}\oplus \bar{u1}$ should describe such a spin-1 boson condensed
state where the charge operators and spin operators both have algebraic
correlations.


The low energy theory of the critical state is completely encapsulated in the partition function. For a CFT to be a bosonic low energy theory, its partition function must be modular invariant. \cite{C8686} To impose the modular invariance condition, it is convenient to consider the
theory on the space-time torus, specified by two independent lattice vector
$\omega_1$ and $\omega_2$ on a complex plane. The partition function of a
conformal invariant theory depends solely on $\tau=\omega_2/\omega_1$, assuming
all the low energy modes, including right and left movers, has the same
velocity that is set to be $1$.  In the Hamiltonian formalism, the partition
function for a bosonic system is given by \eqn{Ztau}. We can use the $u1_N$ and
$su2_k$ characters to construct the modular invariant partition function.
Before the construction, that low energy local operators must be only
flavor-singlet further constrains the allowed operators. In fact, we assume
that {\it all flavor-singlet excitations remain gapless and to carry the
same quantum numbers.} In particular, one allowed type is given by
\begin{align}
 \begin{split}
\epsilon^{vwst}\psi_{v}\psi_{w}\psi_s\psi_t \sim & \ee^{\ii 2\vphi_c} ,\\
\epsilon^{vwst}\bpsi_{v}\bpsi_{w}\bpsi_s\bpsi_t \sim & \ee^{\ii 2\bar\vphi_c} .
\end{split}
\label{su2u1op1}
\end{align}
where $\epsilon^{vwst}$ is the Levi-Civita symbol, with $v,w, s,t$ taking
values in $\{(a,\alpha)|a,\alpha=1,2\}$. They are spin-0 charge-$4$ momentum
$4k_F=k_B$ local bosonic operators with scaling dimensions $(h,\bar h)=(2,0)$
and $(h,\bar h)=(0,2)$. They are also purely chiral operators as they contain
only right movers or only left movers. Next we examine the charge-$2$ spin-0
local bosonic operators
\begin{align}
\label{charge2}
 \psi_{\al a} (\ii\si^2)_{\al\bt} (\ii\si^2)_{ab} \psi_{\bt b}
,\quad \bar \psi_{\al a} (\ii\si^2)_{\al\bt} (\ii\si^2)_{ab} \bar\psi_{\bt b}.
\end{align}
However, in the low energy sector, the only purely chiral charge-$2$ spin-$0$
operators built from \eqn{fer1} are $\ee^{\ii \vphi_c}$ and $\ee^{\ii \bar \vphi_c}$.  They have
scaling dimensions $(h,\bar h)=(\frac12,0)$ and $(h,\bar h)=(0,\frac12)$ with half-integral conformal spin $h-\bar h$  and
are not local bosonic operators.  This implies that the charge-$2$ spin-0 local
bosonic operators in \eqn{charge2} do not belong to the gapless sector, and are
gapped operators.  

Another type of purely chiral low energy local operators is given by
\begin{align}
\label{charge2spin1}
\begin{split}
 \psi_{\al a} (\ii\si^2 \si^l)_{\al\bt} (\ii\si^2)_{ab} \psi_{\bt b} \sim & \ee^{\ii \vphi_c} V^{su2_2^s}_{1,l},\\
 \bar \psi_{\al a} (\ii\si^2 \si^l)_{\al\bt} (\ii\si^2)_{ab} \bar\psi_{\bt b} \sim & \ee^{\ii \bar\vphi_c} \bar V^{su2_2^s}_{1,l}.
\end{split}
\end{align}
These charge-$2$ spin-1 momentum $2k_F=\frac12 k_B$ local bosonic operators are
electron pairs with same crystal momenta, forming singlet in flavor channel and
triplet in spin channel.  They have scaling dimensions $(h,\bar h)=(1,0)$ and
$(h,\bar h)=(0,1)$ also belong to the gapless sector (see Table \ref{za}). 

Since the single chiral fermion contains $\ee^{\ii \varphi_c/2}$ from \eqn{u1Op}, we deduce that the smallest possible radius and level are $R=2$ and $N=4$,
and the $u1$ CFT is more precisely $u1_4$ CFT.

Now we construct the partition functions for the critical state. The partition function must contain a term $ \chi^{u1_4}_0 \chi^{su2_2}_0
\bchi^{u1_4}_0 \bchi^{su2_2}_0 $, which corresponds to the presence of a vacuum
state.   
It must also contain $ \chi^{u1_4}_2 \chi^{su2_2}_1 \bchi^{u1_4}_0
\bchi^{su2_2}_0 $, $ \chi^{u1_4}_0 \chi^{su2_2}_0 \bchi^{u1_4}_2
\bchi^{su2_2}_1 $ and $ \chi^{u1_4}_2 \chi^{su2_2}_1
\bchi^{u1_4}_2 \bchi^{su2_2}_1 $, corresponding to the presence of the local operators in \eq{charge2spin1} and their product. 
Starting from these spin-integer bosonic terms
\begin{align}
(\chi^{u1_4}_0\chi^{su2_2}_0+\chi^{u1_4}_2\chi^{su2_2}_1)(\bchi^{u1_4}_0\bchi^{su2_2}_0+\bchi^{u1_4}_2\bchi^{su2_2}_1)
\end{align}
there is a \emph{single solution} of modular invariant partition function 
\eq{Z1tau}

Such a result implies that the bosonic $Z_A$-CFT can be realized by a spin-$\fh$
fermion system.  And all fermions are gapped since the $Z_A$-CFT contains no fermionic operators. In other words, the $Z_A$-CFT can be realized by
pairing fermions to form spin-1 bosons.  Thus the two spin-gapped superfluids
for spin-1 bosons and their phase transition $su2_2\oplus u1 \oplus
\bar{su2}_2\oplus \bar{u1}$ can be realized, for example, by spin-$\fh$ electrons
on a ladder (with doping).  Such a critical point and its neighboring
spin-gapped states on the ladder has been studied in
\Ref{BFc9503045,LFc9801285,CTc0503050,T181101653}.  We emphasize here the full partition function of the $su2_2\oplus u1
\oplus \bar{su2}_2\oplus \bar{u1}$ critical point, which allows us to calculate
the scaling dimensions and the symmetry properties of \emph{all} the local
operators. This calculation goes beyond the renormalization group calculation
of a few local operators, and can handle more complicated problems where
perturbation calculations may fail. 

The single relevant operator $V_1^{su2_2^s}\bar{V}_1^{su2_2^s}$ written in terms of fermion operators as in (\ref{fperb}) with exchanged the spin and flavor indices.

We have shown a construction of bosonic critical theory from interacting fermion models. However, the construction is still in terms of field theories. In the corresponding lattice model, the interaction term, in particular, would involve complicated form.

\begin{table*}[t]
 \centering
 \def\arraystretch{1.9}
\begin{tabular}
{| c | c | c | c | c | c|}
 \hline
 operators & spin & $q$ & $\frac{q_t}{k_B}$ & $h,\bar h$ & restriction \\
 \hline\hline
 $V_{1,l}\bar V_{1,l'}$ & $0,1,2$ & $0$ & $0$ & $\frac{1}{2},\frac{1}{2}$ & \\
 \hline
  $\ee^{\pm \ii \frac{\sqrt{M}}{2}(\varphi+\bar\varphi)} $ & $0$ & $\pm M$ & $0$ & $\frac{M}{8},\frac{M}{8}$& \\
 \hline
   $\ee^{\pm \ii \frac{\sqrt{M}}{2}(\varphi-\bar\varphi)} $ & $0$ & $0$ & $1$ & $\frac{M}{8},\frac{M}{8}$& \\
\hline
   $\ee^{\pm \ii \frac{m}{\sqrt{M}}(\varphi+\bar\varphi)} $ & $0$ & $\pm  2m$ & $0$ & $\frac{m^2}{2M},\frac{m^2}{2M}$ &  $m=0 \mod 2$ \\
 \hline
   $\ee^{\pm \ii \frac{m}{\sqrt{M}}(\varphi-\bar\varphi)} $ & $0$ & $0$ & $\pm \frac{2m}{M}$ & $\frac{m^2}{2M},\frac{m^2}{2M}$ & $m=0\mod 2$\\
   \hline
      $\ee^{\pm \ii \frac{m}{\sqrt{M}}(\varphi+\bar\varphi)}V_{\frac{1}{2},l}\bar V_{\frac{1}{2},l'} $ & $0,1$ & $\pm 2m$ & $0$ & $\frac{m^2}{2M}+\frac{3}{16} ,\frac{m^2}{2M}+\frac{3}{16}$ & $m=1 \mod 2$ \\
 \hline
   $\ee^{\pm \ii \frac{m}{\sqrt{M}}(\varphi-\bar\varphi)}V_{\frac{1}{2},l}\bar V_{\frac{1}{2},l'}  $ & $0,1$ & $0$ & $\pm \frac{2m}{M}$ & $\frac{m^2}{2M}+\frac{3}{16},\frac{m^2}{2M}+\frac{3}{16}$ & $m=1 \mod 2$ \\
   \hline\hline
$\ee^{\pm \ii \left[\left(\frac{\sqrt{M}}{2}+\frac{m}{\sqrt{M}}\right)\varphi +\frac{m}{\sqrt{M}}\bar \varphi\right]} V_{1,l}$ & $1$ & $\pm \left(\frac{M}{2}+2m\right)$ & $\pm \frac{1}{2}	$ & $\frac{\left(m+\frac{M}{2}\right)^2}{2M}+\frac{1}{2},\frac{m^2}{2M}$ & $m- \frac{M}{4}=1 \mod 2$ \\
\hline
$\ee^{\pm \ii \left[\left(\frac{\sqrt{M}}{2}+\frac{m}{\sqrt{M}}\right)\varphi +\frac{m}{\sqrt{M}}\bar \varphi\right]} \bar V_{1,l}$ & $1$ & $\pm \left(\frac{M}{2}+2m\right)$ & $\pm \frac{1}{2}	$ & $\frac{\left(m+\frac{M}{2}\right)^2}{2M},\frac{m^2}{2M}+\frac{1}{2}$ & $m- \frac{M}{4}=1 \mod 2$  \\
\hline
$\ee^{\pm \ii \left[\left(\frac{\sqrt{M}}{2}+\frac{m}{\sqrt{M}}\right)\varphi +\frac{m}{\sqrt{M}}\bar \varphi\right]} V_{\frac{1}{2},l}\bar V_{\frac{1}{2},l'}$ & $0,1$ & $\pm \left(\frac{M}{2}+2m\right)$ & $\pm \frac{1}{2}	$ & $\frac{\left(m+\frac{M}{2}\right)^2}{2M}+\frac{3}{16},\frac{m^2}{2M}+\frac{3}{16}$ & $m- \frac{M}{4}=0 \mod 2$  \\
\hline\hline
$\ee^{\pm \ii \left[\left(\frac{\sqrt{M}}{2}+\frac{m}{\sqrt{M}}\right)\bar \varphi +\frac{m}{\sqrt{M}} \varphi\right]} V_{1,l}$ & $1$ & $\pm \left(\frac{M}{2}+2m\right)$ & $\mp \frac{1}{2}	$ & $\frac{m^2}{2M} +\frac{1}{2},\frac{\left(m+\frac{M}{2}\right)^2}{2M}$ & $m- \frac{M}{4}=1 \mod 2$ \\
\hline
$\ee^{\pm \ii \left[\left(\frac{\sqrt{M}}{2}+\frac{m}{\sqrt{M}}\right)\bar \varphi +\frac{m}{\sqrt{M}} \varphi\right]} \bar V_{1,l}$ & $1$ & $\pm \left(\frac{M}{2}+2m\right)$ & $\mp \frac{1}{2}	$ & $\frac{m^2}{2M},\frac{\left(m+\frac{M}{2}\right)^2}{2M}+\frac{1}{2}$ & $m- \frac{M}{4}=1 \mod 2$  \\
\hline
$\ee^{\pm \ii \left[\left(\frac{\sqrt{M}}{2}+\frac{m}{\sqrt{M}}\right)\bar \varphi +\frac{m}{\sqrt{M}}\varphi\right]} V_{\frac{1}{2},l}\bar V_{\frac{1}{2},l'}$ & $0,1$ & $\pm \left(\frac{M}{2}+2m\right)$ & $\mp \frac{1}{2}	$ & $\frac{\left(m+\frac{M}{2}\right)^2}{2M}+\frac{3}{16},\frac{m^2}{2M}+\frac{3}{16}$ & $m- \frac{M}{4}=0 \mod 2$  \\
\hline
 \end{tabular}
 \caption{$\ZZ_2$ orbifold $u(1)_M\oplus su(2)_2$ with $M=0\mod 4$ critical point. $m$ is an integer in $[0,\frac{M}{2})$. The distinction for $M=0 \mod 8$ and $M=4 \mod 8$ cases is shown in the restrictions. The top block contains non-chiral/untwisted operators and the bottom two blocks contain chiral/twisted operators. }
 \label{M4}
 \end{table*}

\section{$\ZZ_2$ orbifold of $U(1)\times SU(2)$ CFTs}\label{z2orbifold}
\subsection{$\ZZ_2$ orbifold construction of $u(1)_4\oplus su(2)_2$ CFT}

For concreteness, we start with $U(1)_4\times SU(2)_2$ WZW theory, the fields $g(z,\bar z)$ take values in $(\phi_c(z,\bar z), v(z, \bar z))$, where $\phi_c$ is an angular variable, $\phi_c\cong \phi_c+2\pi R$ is the field in $u(1)_4$ CFT with $R=2$. And $v(z,\bar z)$ are matrix representations of $SU(2)$. Its partition function is $Z_B$, also called diagonal partition function $Z_{\text{diag}}$. The WZW action is invariant under $g_Lg g_R^{-1}$, where $g_L(z), g_R(\bar z)\in U(1)\times SU(2)$. Consider a $\ZZ_2$ subgroup in $U(1)\times SU(2)$.
The nontrivial element in $\ZZ_2$ is the $\pi$ rotation of $U(1)$ part and the $\pi$ rotation around $\hat z$-axis of $SU(2)$. 
\begin{align}
\v a: \left(\phi_c(z,\bar z), v(z,\bar z)\right)\rightarrow \left(\phi_c(z,\bar z)+2\pi\frac{R}{2}, -v(z,\bar z)\right)
\end{align}
For a representation $|k,j\rangle, k=0,1,2,3; j=0,\frac{1}{2},1$, which corresponds to the operator $\ee^{\ii \frac{k}{2}\phi_c} v_{j,\bar j}$, it transforms as
\begin{align}
\v a|k,j\rangle=(-1)^{k+2j}|k,j\rangle
\end{align}



In $Z_{\text{diag}}$-CFT, the field $g(z,\bar z)$ has periodic boundary condition in both space and time direction, 
\begin{align}
\begin{split}
g(z+1, \bar z+1)=& a_1 g(z,\bar z) \\
 g(z+\tau, \bar z+\bar \tau)=& a_0 g(z,\bar z)
\end{split}
\end{align}
where $a_0, a_1$ are identity. 
Now we twist $Z_{\text{diag}}$-CFT by $\ZZ_2$ group. The symmetry twist means to identify the fields differing by an symmetry action in $\ZZ_2$. This generalizes the allowed boundary condition to that $a_0, a_1$ taking elements in $\ZZ_2$. The partition function is a summation of ones with all allowed boundary conditions, 
\begin{align}
Z_{\text{orb}}=\frac{1}{2}\sum_{a_0, a_1\in \ZZ_2} Z_{a_1, a_0}
\end{align}
which is called an orbifold CFT. $Z_{\text{orb}}$ is modular invariant if it satisfies the following 
\begin{align}
\calS Z_{a_1, a_0} = Z_{a_0, a_1},\quad \calT Z_{a_1, a_0}= Z_{a_1, a_1 a_0}
\end{align}

Now we describe a general procedure to construct an orbifold partition function from a diagonal one,\cite{francesco2012conformal} taking $Z_B$ as a concrete example. First, we project the diagonal CFT $Z_{\text{diag}}=Z_B$ shown in (\ref{zb}) where taking $M=4$, to the $\ZZ_2$ symmetric partition function, that is, to keep characters in $Z_{\text{diag}}$ such that $k+2j\in 2\ZZ$, 
\begin{align}
Z_{\text{proj}}=&\left( |\chi_0^{u1_4}|^2+|\chi_2^{u1_4}|^2\right) \left( |\chi_0^{su2_2}|^2+|\chi_1^{su2_2}|^2\right) \nn\\
&+\left( |\chi_1^{u1_4}|^2+|\chi_3^{u1_4}|^2\right) |\chi_{\frac{1}{2}}^{su2_2}|^2
\end{align}

Next, we construct the following linear combination of partition functions, 
\begin{align}
Z_A=(1+\calS+\calT\calS)Z_{\text{proj}}-Z_{\text{diag}}
\end{align}
and find that it is modular invariant. 

In $\ZZ_2$ orbifold CFT, the chiral charge-2 spin-1 operator $\ee^{\ii \varphi}V_{1,l}^{su2_2^s}$ acts like a ``soliton'' (or twist) operator similarly as the spin operator $\sigma$ in the CFT of a free Majorana fermion, which changes the fermion boundary condition from periodic to anti-periodic. Its operator product expansions with operators in the periodic sector $Z_{\text{proj}}$ generate the anti-periodic/twisted sector in $Z_A$.

\subsection{$\ZZ_2$ orbifold $u(1)_M\oplus su(2)_2$  CFT for $M=4 \mod 8$ and $M=0 \mod 8$}

More generally, the orbifold CFTs exist in $su(2)_2\oplus u(1)_M$ for $M=0 \mod 4$. Those with $M=4\mod 8$ has similar structure as the case $M=4$. The soliton field is
\begin{align}
O_{\frac{M}{2},1}=\ee^{\ii \frac{M/2}{\sqrt{M}}\varphi} V_{1,l}^{su2_2^s}
\end{align}
with scaling dimension $(h,\bar h)= \left(\frac{M}{8}+\frac{1}{2},0\right)$ and charge-$\frac{M}{2}$ spin-$1$. Since $h$ is integral, the twist field is also a bosonic field. 
The partition function is 
\begin{align}
Z^{M4}_{\text{orb}}= &\left(\sum_{0\leq j<M, \text{even } j} |\chi^{u1_M}_{j}|^2\right)\left(|\chi_0^{su2_2^s}|^2+|\chi_1^{su2_2^s}|^2\right)\nn\\
&+\left(\sum_{0\leq j<M, \text{odd } j} |\chi^{u1_M}_{j}|^2\right)|\chi_{\frac{1}{2}}^{su2_2^s}|^2\nn\\
&+\sum_{0\leq j<M, \text{even } j} \left(\chi^{u1_M}_{j}\chi^{su2_2^s}_{0}\bar{\chi}^{u1_M}_{j+M/2 \mod M}\bar{\chi}^{su2_2^s}_{1}\right.\nn\\
&+ \left.\chi^{u1_M}_{j}\chi^{su2_2^s}_{1}\bar{\chi}^{u1_M}_{j+M/2 \mod M}\bar{\chi}^{su2_2^s}_{0}\right)\nn\\
&+\sum_{0\leq j<M, \text{odd } j} \left(\chi^{u1_M}_{j}\bar{\chi}^{u1_M}_{j+M/2 \mod M}|\chi^{su2_2^s}_{\frac{1}{2}}|^2\right)
\end{align}

The operator content is listed in Table \ref{M4}. And there is one operator carrying charge-2 spin-1.

In the case $M=0\mod 8$, the scaling dimension $h$ of the chiral operator $O_{\frac{M}{2},1}$ is half-integral. Therefore, it is forbidden in a modular invariant partition function. Nevertheless, the orbifold partition function is
\begin{align}
Z_{\text{orb}}^{M8}=&\left(\sum_{0\leq j<M, \text{even } j} |\chi^{u1_M}_{j}|^2\right)\left(|\chi_0^{su2_2^s}|^2+|\chi_1^{su2_2^s}|^2\right)\nn\\
&+\left(\sum_{0\leq j<M, \text{odd } j} |\chi^{u1_M}_{j}|^2\right)|\chi_{\frac{1}{2}}^{su2_2^s}|^2\nn\\
&+\sum_{0\leq j<M, \text{odd } j} \left(\chi^{u1_M}_{j}\chi^{su2_2^s}_{0}\bar{\chi}^{u1_M}_{j+M/2 \mod M}\bar{\chi}^{su2_2^s}_{1}\right.\nn\\
&+ \left.\chi^{u1_M}_{j}\chi^{su2_2^s}_{1}\bar{\chi}^{u1_M}_{j+M/2 \mod M}\bar{\chi}^{su2_2^s}_{0}\right)\nn\\
&+\sum_{0\leq j<M, \text{even } j} \left(\chi^{u1_M}_{j}\bar{\chi}^{u1_M}_{j+M/2 \mod M}|\chi^{su2_2^s}_{\frac{1}{2}}|^2\right)
\end{align}

Here, the twist operator is 

\begin{align}
\ee^{\ii \frac{M/2}{\sqrt{M}}\varphi}V_{\frac{1}{2},l}^{su2_2^s}\bar{V}_{\frac{1}{2},l'}^{su2_2^s},
\end{align}
with scaling dimension $(h,\bar h)= \left(\frac{M}{8}+ \frac{3}{16}, \frac{3}{16}\right)$, which is no longer a chiral current. 

The operator content is also listed as in Table \ref{M4}, but with different restrictions (last column of the table). We see all orbifold models have charge-2 spin-1 bosonic excitations. Each has a single relevant operator with trivial quantum numbers. They may all be effective theories to describe the critical points between TSF ad nTSF. How to distinguish them physically requires further studies.

\bibliography{SFspin1_arXiv,wencross,all,publst}


\end{document}